\documentclass{lmcs} 
\usepackage[utf8]{inputenc}
\usepackage{graphicx}
\usepackage{hyperref}
\keywords{Android, logic circuit, simulation, mobile application, Unity.}

\usepackage{hyperref}

\theoremstyle{plain} 

\begin{document}

\title[Simulation of Logic Circuit Tests on Android-Based Mobile Devices]{Simulation of Logic Circuit Tests on Android-Based Mobile Devices}

\author[A.~Çakir]{ABDÜLKADİR ÇAKIR}	
\address{Department of Electrical and Electronics Engineering, Faculty of Technology, Suleyman Demirel University, Isparta, Turkey}	
\email{abdulkadircakir@sdu.edu.tr, ummusanc32@gmail.com}  

\author[Ü.~Çıtak]{ÜMMÜŞAN ÇITAK}

\begin{abstract}
  \noindent In this study, an application that can run on Android and Windows-based mobile devices was developed to allow students attending such classes as Numerical/Digital Electronics, Logic Circuits, Basic Electronics Measurement and Electronic Systems in Turkey’s Vocation and Technical Education Schools to easily carry out the simulation of logic gates, as well as logic circuit tests performed using logic gates. A 2D-mobile application that runs on both platforms was developed using the C\# language on the Unity3D editor.
  
To assess the usability of the mobile application, a one-hour training session was administered in March of the 2017-2018 academic year to two groups of students from a single class in the sixth grade of an Imam Hatip Secondary School affiliated to the Ministry of National Education. Each of the two groups contained 12 students who were assumed to be equivalent, and who had no prior knowledge of the subject. The training of the first group began with a lecture on basic logic gates using a blackboard, and involved no simulations, while the second group, in addition to being given the same the lecture, received additional training involving demonstrations of the developed mobile application and its simulations. Following the lectures, a written exam was applied to both groups. An evaluation of the exam results revealed that 83 percent of the students who had been given demonstrations of the mobile application were able to perform the circuit task completely, whereas only 50 percent of the other were able to complete the task. It was concluded that the application was both useful and facilitating for to the students, and it was also noted that students who were supported by the mobile application had gained a better grasp of the topic by being able to see and practice the simulations first hand.

\end{abstract}

\maketitle

\section{Introduction}\label{S:one}

  Education occupies an important place not only in the school life of individuals, but also in every aspect of daily life. Recent technological developments have led to the emergence of countless new learning environments for individuals, with one such environment being the medium provided by mobile devices and tablet computers, which can be accessed independent of location and time. These technologies are becoming more indispensable with each passing day, and the interest of students in different courses, as well as their levels of achievement, can be increased by rendering abstract concepts more tangible through the use in classes of the mobile simulation applications that have become widespread in recent years. Mobile learning software can also reduce the challenges experienced in real applications, and can circumvent the potential risks to which users may be exposed, while also allowing learning whenever needed, with no constraints of location or time. 

  In line with the curriculum of the course entitled Digital Circuits 1 given in Marmara University’s Vocational School of Technical Sciences, Altıkardeş (2001) prepared a web-based education material application entitled Logic Circuits 1 that allowed students to follow the aforementioned course via the Internet, in support of classical education methods \cite{alt}.
  
  In another study, Ekiz et al. (2003) described the preparation of a set of criteria for to enable the remote teaching of the course entitled Logic Circuits, or Digital Electronics, making use of the an Internet Supported Education (ISE) approach. To this end, they introduced the necessary tools/methods for the creation/development the of course content, while also making an assessment of Internet-supported distance education \cite{ebu}. 
  
  The study of Büyükbayraktar (2006) aimed at determining the effects of the Proteus virtual laboratory program, developed with the aim of simulating analog and digital electronic circuits, on the performance of students studying Logic Circuits and those involved in the Digital Electronics workshop \cite{buy}.
  
  Oran and Karadeniz (2007) made a theoretical discussion of the role of mobile learning in Internet-based education, highlighting its advantages as well as the problems that need to be overcome \cite{ok}.
  
  In their study, Rosilah et al. (2008, 2010) proposed a practical new laboratory practice and experience approach to the Digital Logic Design course that can be accessed via the Internet \cite{rnhs, rnhs10}.
  
  Çakır (2011) developed a mobile software for a Basic Information Technology course, and made an assessment of student opinion related to the software \cite{cak}.
  
  In his study, Cupic (2011) developed a web technology-based integrated platform for the teaching and learning of Digital Logic \cite{cup}.
  
  Akkağıt and Tekin (2011) meanwhile, developed an educational tool for logic gates using LabVIEW software \cite{at}.
  
  Using LabVIEW 2010 Student Edition software, Bal (2012) designed a toolbox entitled Digital Logic that students from in the departments of Electric-Electronics Engineering, Computer Engineering and Computer Sciences could use in applications related to the Digital Logic Design course \cite{bal}.
  
  Yechshzhanova (2014) developed the content of a geometry course for CAE and Mobile Learning, using 3DS MAX and UNITY3D software in for development of the content \cite{yec}.
  
  In his study, Dehmenoğlu (2015) developed an Android-based mobile application entitled Code Everywhere that offers support to those taking the Basics of Programming course in the Departments of Information Technologies in Vocational and Technical Anatolian High Schools \cite{deh}.
  
  To investigate the impact of mobile technologies on student performance, Kalınkara (2017) used the Android operating system platform to develop an application covering the subjects in the Computer Hardware and Electronics course \cite{kal}.
  
  With each passing day, android-based smart devices are gaining more significance in our daily lives, and foremost among these devices are smart phones and tablet PCs \cite{set}. Over time, there has been a growth in the number of projects carried out that take advantage of the superior features of these devices, which greatly facilitate human life. With the prevalence of these portable systems, people are today able to access many applications without being hindered by the constraints of time or space.
  
  The rapid development of technologies has also had a profound effect on education. The use of smart boards in educational institutions, the provision of tablet computers to students, and the ability to follow lessons via the Internet are just some examples of the close relationship between technological development and educational tools \cite{dybc}. 
  
  In recent years, there have been a variety of mobile applications and academic studies developed in many fields based on simulation-based learning. A review of literature uncovers several android-based mobile applications related to the use of Logic Gates and Logic Circuits that are available from Google Play. That said, it is apparent that while academic studies conducted into this subject have addressed the development of computer-aided simulation applications, there have to date been no mobile simulation applications developed that are capable of running on both Android and Windows platforms. 
  
  In the present study, an application is developed that can run on Android- and Windows-based mobile devices. A 2D application was developed in C\# on the Unity3D editor that allows students to carry out simulations of logic circuit tests, prepared using logic gates either during class or at a time and place of their choice, and which are also capable of operating on the Tablet PCs distributed in schools within the scope of the FATİH Project (Project to Increase Opportunities and Improve Technology), on personal smartphones, and on Windows-based smart boards and touchscreen laptops.
  
\section{Material Method}

  The study instructs students on how to carry out in-class simulations of logic circuit tests on mobile devices for such courses such as Numerical/Digital Electronics, Basic Electronics and Measurement, and Electronic Systems, all of which are provided in Vocation and Technical Education schools.
  
  Prior to developing an application for Android in the Unity3D Game Engine, it was first necessary to install a Java JDK (Java Development Kit) onto the computer, and afterwards, an Android SDK (Software Development Kit) was installed using the Android SDK Manager. When installing the Android SDK, a version superior to the Android 2.2 was chosen, and all installation options were selected as necessary. Following this, the latest version of the Unity3D software was installed after downloading it free of charge from the website. When installing Unity3D, the Visual Studio 2015 software used for writing in C\# code was also installed.
  
\subsection{Software of the Used Simulation}
  
\begin{figure}
\begin{center}
\includegraphics[width=9.cm, height=5.cm]{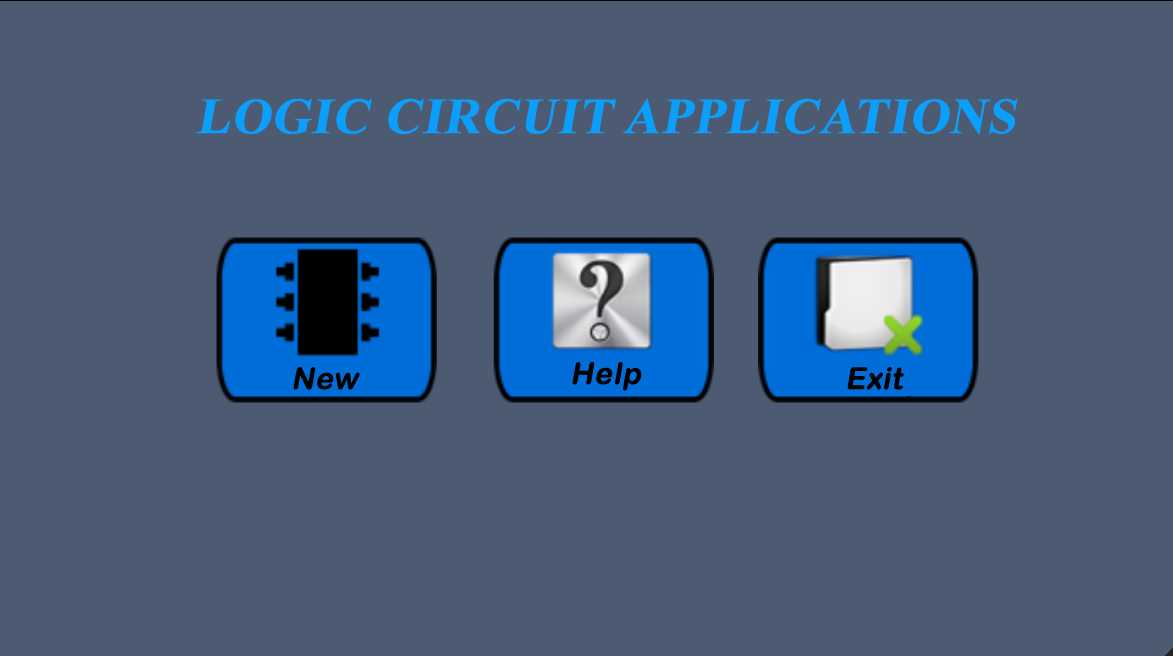}
\end{center}
\caption{Home screen view }
\label{fig:1}
\end{figure}

  The mobile application comprises two parts: the Home Screen and the Simulation Screen. The Home Screen contains New, Help and Exit buttons, as can be seen in Fig. \ref{fig:1}. 

\begin{figure}
\begin{center}
\includegraphics[width=9.cm, height=5.cm]{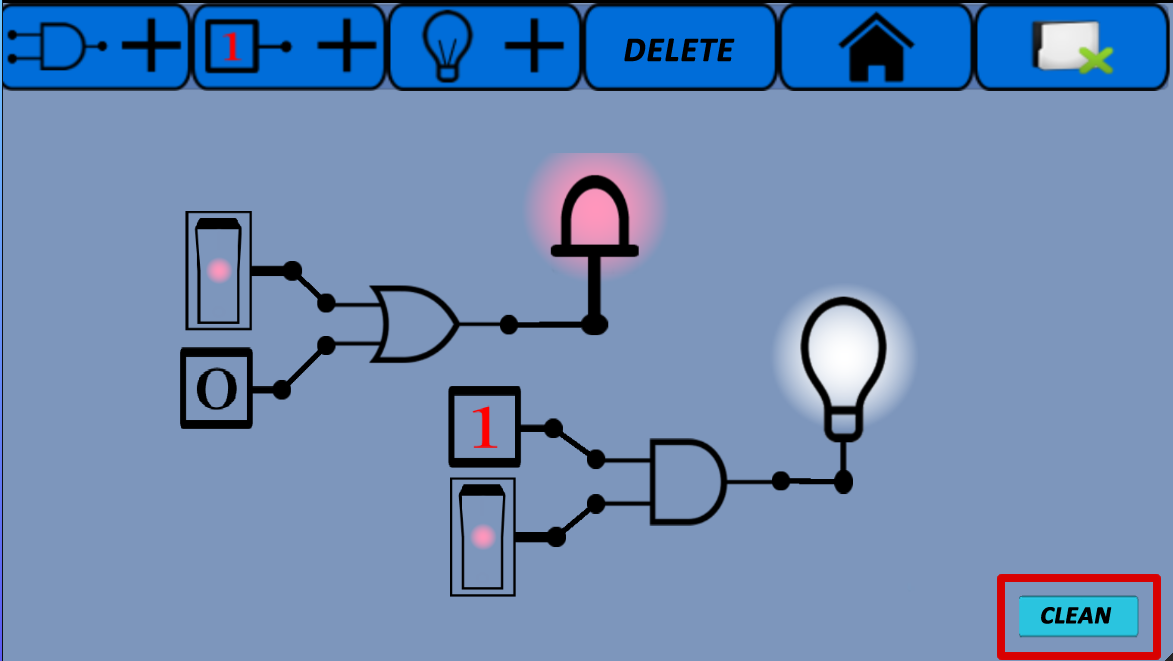}
\end{center}
\caption{Simulation screen view }
\label{fig:2}
\end{figure}

The Exit button on the screen is used to completely exit the software, while tapping the New button opens the Simulation screen, where a new circuit application can made. Fig. \ref{fig:2} shows the Simulation screen on which the circuit applications are carried out. A small code ensures the screen remains horizontal during the operation of the software, facilitating use.  

\begin{figure}
\begin{center}
\includegraphics[width=9.cm, height=5.cm]{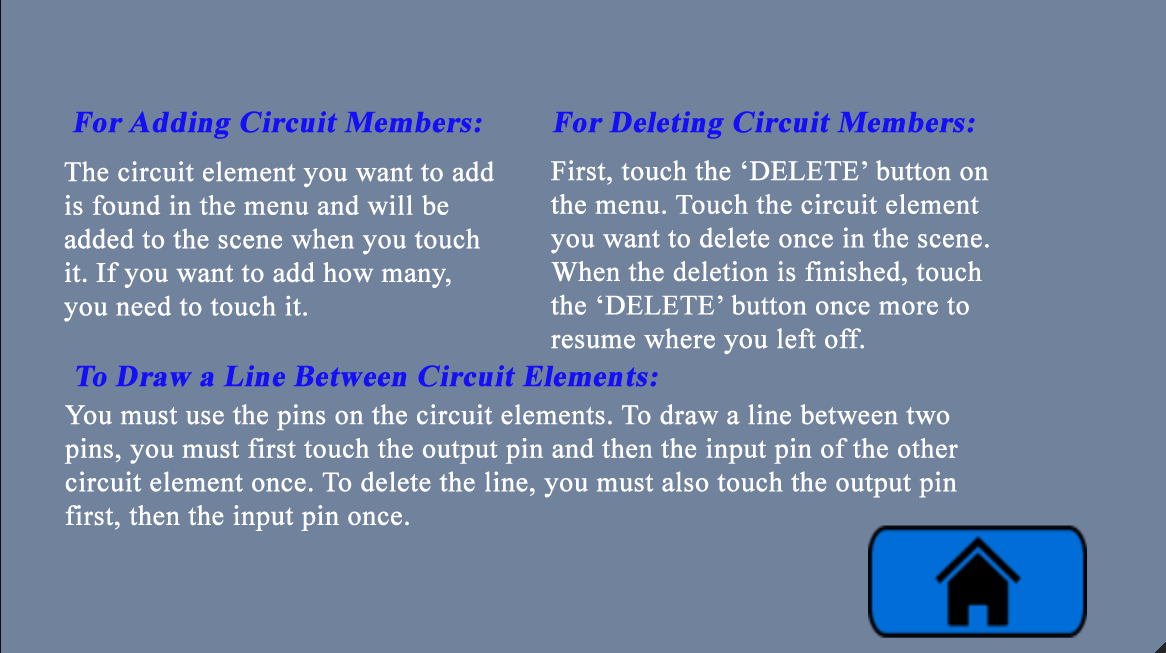}
\end{center}
\caption{Help screen view  }
\label{fig:3}
\end{figure}

Tapping the Help button opens the Help screen, where an explanation is given of how the software operates. The Help screen shown in Fig. \ref{fig:3} describes how to add or delete circuit elements to or from the screen, and on how to create circuits. The button at the bottom of the screen is used to return to the Home screen.

\subsubsection{Simulation screen}
\begin{figure}
\begin{center}
\includegraphics[width=14.cm, height=1.cm]{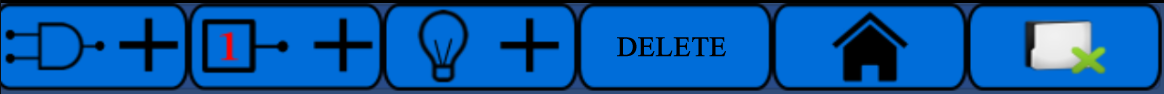}
\end{center}
\caption{Simulation screen menu view}
\label{fig:4}
\end{figure}

  The Simulation screen is where the applications will be performed, and contains the menu from which the necessary circuit elements can be selected. The desired circuit elements are selected from the drop-down list and then added to the circuit. From the menu, it is possible to perform the Add Logic Gate, Add Logic Entry, Add Exit (lamp or led), Delete, Go to Main Menu (Home Screen), and Exit functions. Fig. \ref{fig:4} shows the layout of the menu.
  
\begin{enumerate}
\item\textbf{Add Logic Gate Button:} Tapping the Add Logic Gate button, which is the first option on the menu, opens a list of the gates that can be selected and added to the screen. When any one of the listed gates is tapped, the selected gate appears in the center of the screen. Fig. \ref{fig:5} shows the drop-down list of the Add Logic Gate button.
  
\begin{figure}
\begin{center}
\includegraphics[width=9.cm, height=5.cm]{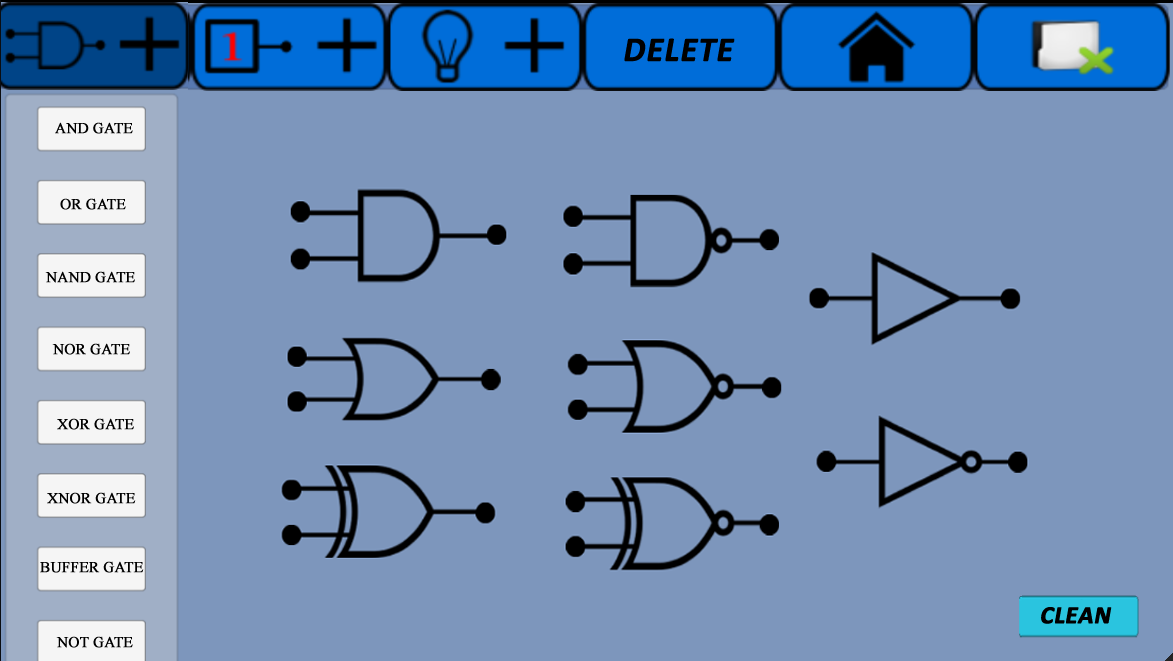}
\end{center}
\caption{Add Logic Gate button}
\label{fig:5}
\end{figure}
  
\item\textbf{Add Logic Input (Load) Button:} Tapping the Add Logic Input button which is the second option on the menu, opens a list showing the inputs that can be selected and added to the screen. The three available options are Logic 1, Logic 0 and Switch. Fig. \ref{fig:6} shows the list that opens when the button is tapped.

\begin{figure}
\begin{center}
\includegraphics[width=10.cm, height=5.cm]{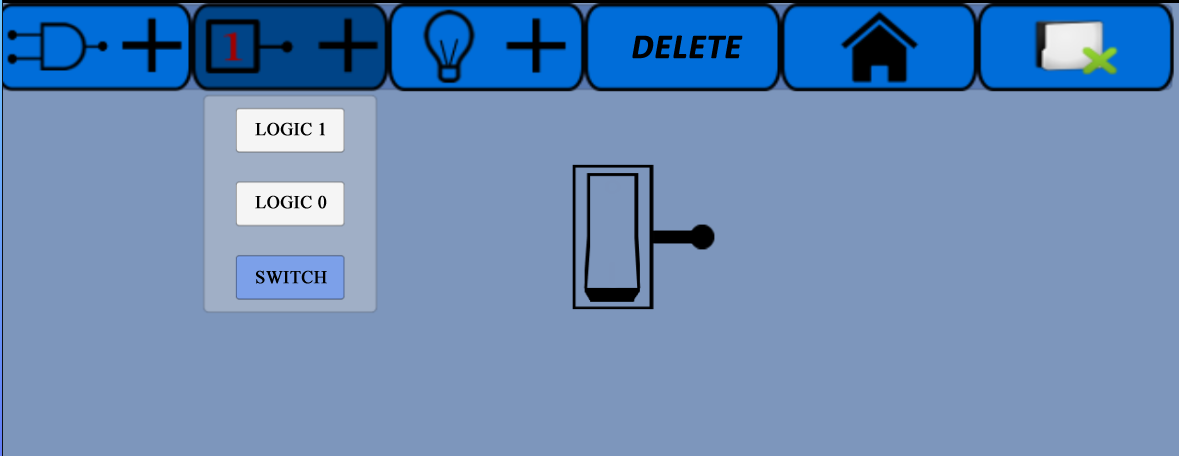}
\end{center}
\caption{Add Logic Input button}
\label{fig:6}
\end{figure}
  
\item\textbf{Add Exit (lamp or led) Button:} Tapping the Add Exit button, which is the third option on the menu, opens a list of the available exits (lamp or led) that can be added to the screen. Operation of the circuit is ensured by having the lamp or led remain on or off, depending on the Logic 1 or Logic 0 arriving at the outputs of the circuit relative to the inputs. Fig. \ref{fig:7} shows the drop-down menu that opens after tapping the Add Exit button.

\begin{figure}
\begin{center}
\includegraphics[width=10.cm, height=5.cm]{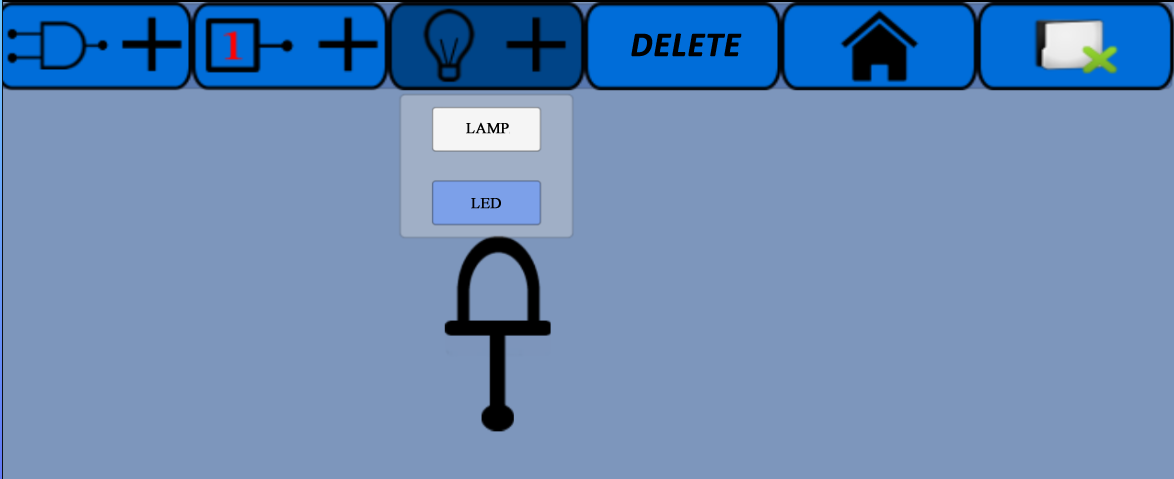}
\end{center}
\caption{Add Exit button}
\label{fig:7}
\end{figure}
  
\item\textbf{Delete Button:} The Delete button on the menu is used to delete circuit elements. Tapping the Delete button causes the name of the button to change to “DELETE ACTIVE”, and the circuit element on the screen can be deleted by tapping the circuit element selected for deletion. All other elements intended for deletion can be removed using the same method. To continue setting up the circuit once the delete process was has been completed, tap the Delete Active button on the menu to change its status back to “DELETE”, after which the editing process can be continued. Fig. \ref{fig:8} shows both the passive and active statuses of the Delete button.

\begin{figure}
\begin{center}
\includegraphics[width=2.cm, height=1.cm]{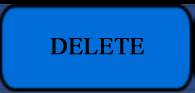}
\includegraphics[width=2.cm, height=1.cm]{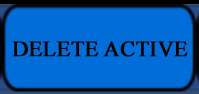}
\end{center}
\caption{The Delete button}
\label{fig:8}
\end{figure}
  
\item\textbf{Go to Main Menu Button:} When the Go to Main Menu button shown in Fig. \ref{fig:9} is tapped on the relevant menu, a pop-up window opens on the screen asking for confirmation whether the user wants to return to the Main Menu (Home Screen). As shown in Fig. \ref{fig:10}, touching the Yes button on the screen returns the user to the Home Screen, while touching the No button returns the user to the screen on which the circuit is being constructed.

\begin{figure}
\begin{center}
\includegraphics[width=2.cm, height=1.cm]{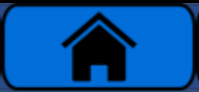}
\end{center}
\caption{Go to Main Menu button}
\label{fig:9}
\end{figure}

\begin{figure}
\begin{center}
\includegraphics[width=9.cm, height=5.cm]{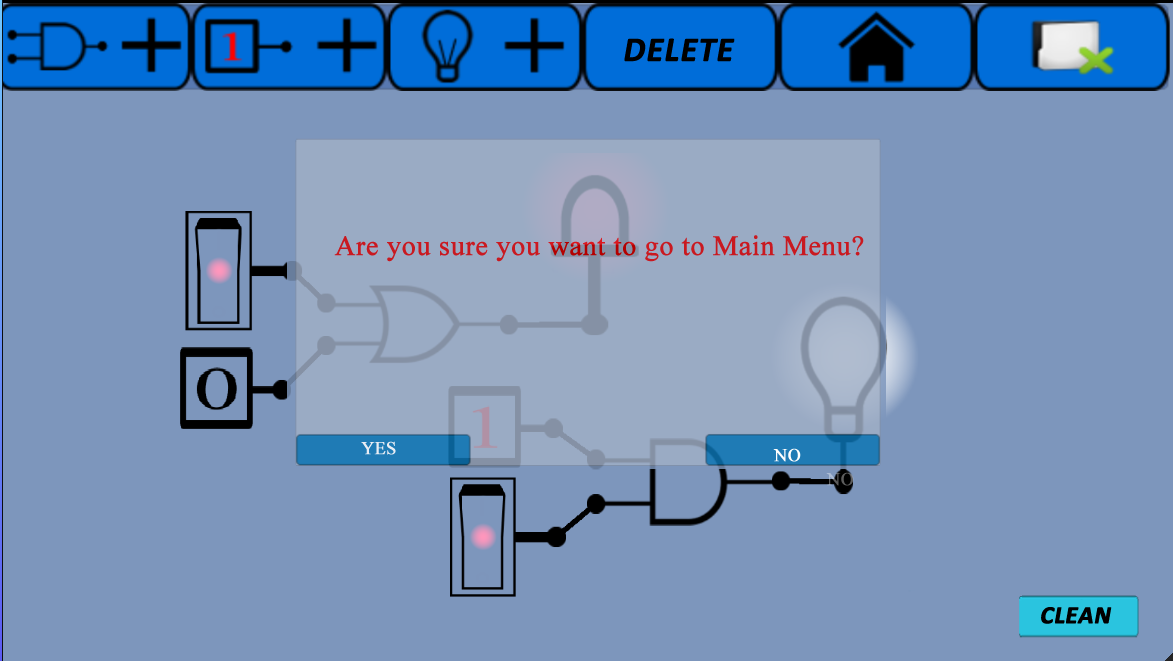}
\end{center}
\caption{Pop-up window that appears when the Go to Main Menu button is tapped}
\label{fig:10}
\end{figure}
 
\item\textbf{Exit Button:} Tapping the Exit button shown in Fig. \ref{fig:11} opens up a pop-up window on the screen asking for confirmation that the user wants to exit the program. As shown in Fig. \ref{fig:12}, the window in question closes if the Yes button is tapped, while touching tapping the No button returns the user to the screen on which the circuit is being constructed.

\begin{figure}
\begin{center}
\includegraphics[width=2.cm, height=1.cm]{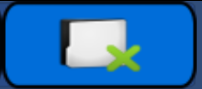}
\end{center}
\caption{Exit button}
\label{fig:11}
\end{figure}

\begin{figure}
\begin{center}
\includegraphics[width=9.cm, height=5.cm]{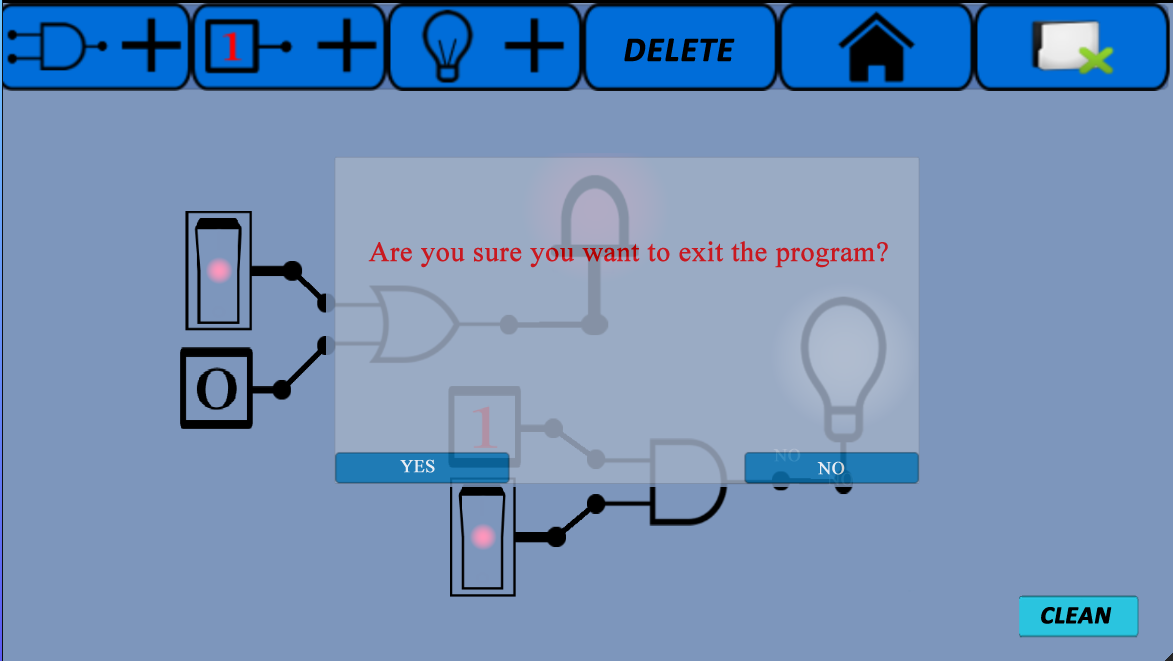}
\end{center}
\caption{View of the window opened by tapping the Exit button}
\label{fig:12}
\end{figure}
  
\item\textbf{Clean Button:} Fig. \ref{fig:13} shows the Clean button, which is located in the lower right-hand corner of the screen, and is used to clear the screen completely, allowing a new circuit to be set up. This should not be confused with the Delete button, was is used to delete specific elements on the screen.

\begin{figure}
\begin{center}
\includegraphics[width=2.cm, height=1.cm]{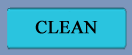}
\end{center}
\caption{Clean button}
\label{fig:13}
\end{figure}
 
\end{enumerate} 
  
\subsection{Flow Diagram of the Simulation Software}
\begin{figure}
\begin{center}
\includegraphics[width=11.cm, height=19.cm]{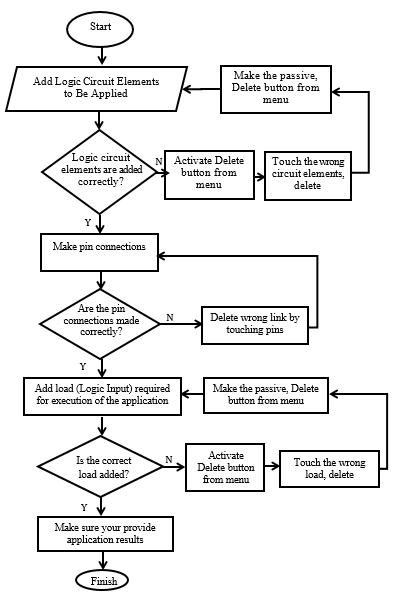}
\end{center}
\caption{Flow diagram showing the operation of the program}
\label{fig:14}
\end{figure}

The circuit elements needed to establish the circuit are selected from the menus and added to the screen. Each element on the screen is moved into position with a “drag and drop” action. If a circuit element is added by mistake, or in excess, it can be deleted from the screen using the Delete button. Finally, connections are made between the pins of the circuit elements, with each connection having to be made one-by-one. To make a connection, first, tap the output pin, followed by the input pin or led pin of the gate to which it is to be connected, taking the direction of the loads into account. The loads are transferred by making connections between the circuit elements. The scripts on the gates that allow them to execute their tasks are added as Components. In these scripts, loads arriving at the inputs of the gates are processed in line with the function of the gate, and the results are transferred to the output pin. Once the connection has been established, the load of the result value is transferred to the next gate or led in line. Fig. \ref{fig:14} presents a flow chart of the running program. 

\subsection{Zooming and Panning on the Screen}
Once the software is prepared, various application examples were carried out. These applications were the And gate practice, the Or gate practice on a smart board, and a half-adder logic circuit.
\subsection{Example of a Simulation Run with the Developed Mobile Application}
While setting up a circuit using the mobile application, it is possible to zoom in or out on of the screen, and to pan up, down, right or left. After zooming in or out and/or completing the circuit, all returning the application to its normal appearance can be achieved by double-tapping an empty area on the screen.
\subsubsection{And Gate Application }
\begin{figure}
\begin{center}
\includegraphics[width=9.cm, height=6.cm]{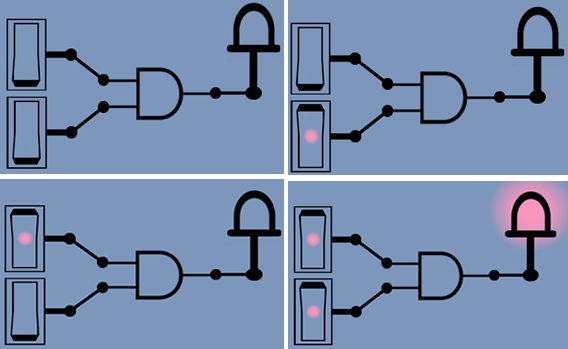}
\end{center}
\caption{And gate application screenshots}
\label{fig:15}
\end{figure}

Fig. \ref{fig:15} shows screenshots of a simulation in which the inputs to the And gate are applied in sequence, with the outputs determined according to the inputs. A Switch was used to apply the input loads.

\subsubsection{Application of an Or Gate On The Smart Board }
\begin{figure}
\begin{center}
\includegraphics[width=11.cm, height=9.cm]{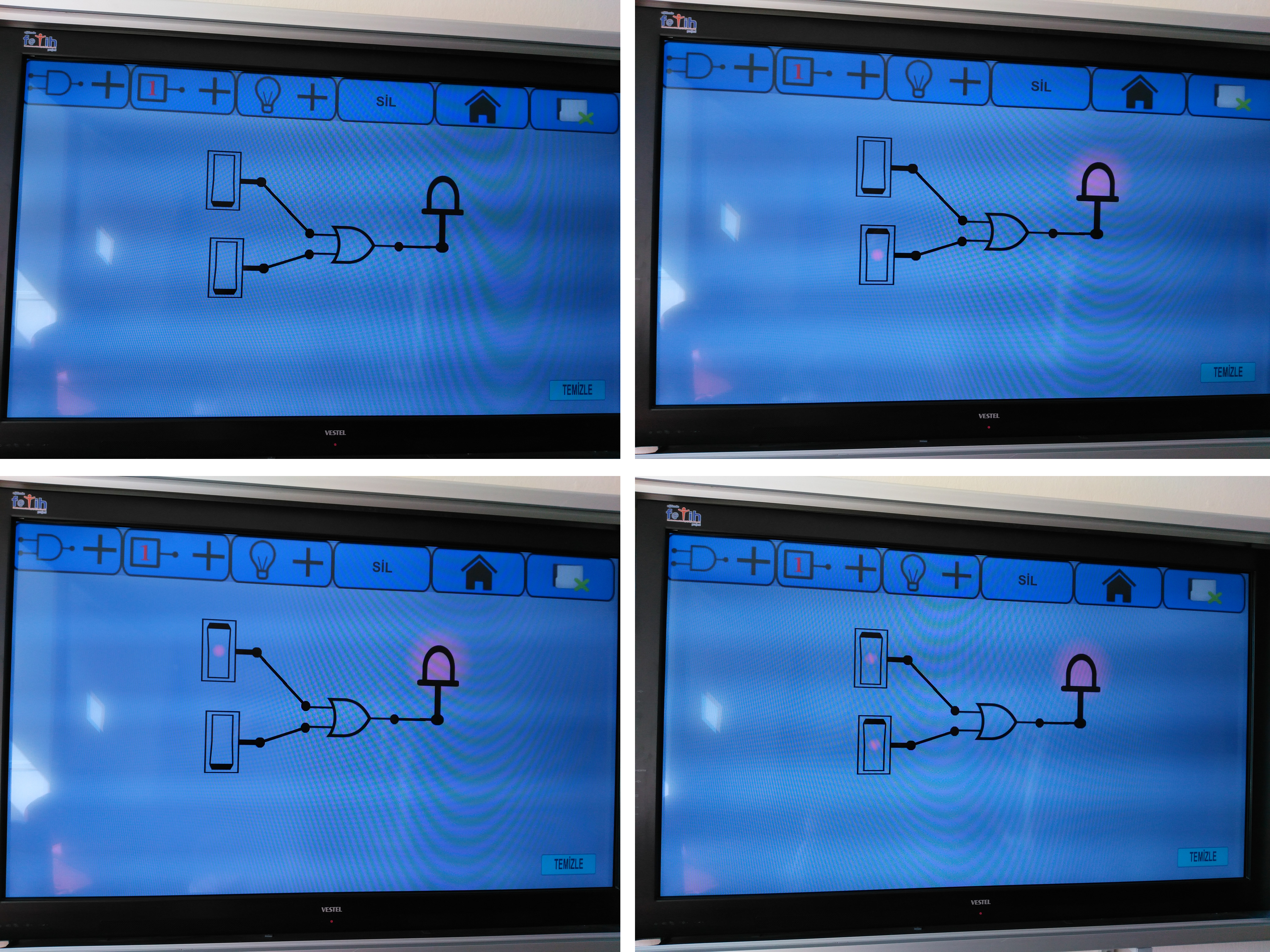}
\end{center}
\caption{Or gate smart board screenshots}
\label{fig:16}
\end{figure}

Fig. \ref{fig:16} shows screenshots of a simulation in which the inputs to an Or gate were applied in sequence, with the outputs determined according to the inputs. The simulation of the Or gate was performed by running the mobile application on a smart board.

\subsubsection{Application Of The Half-Adder Logic Circuit }
\begin{figure}
\begin{center}
\includegraphics[width=6.cm, height=2.cm]{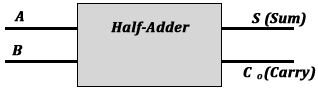}
\end{center}
\caption{Half-adder symbol}
\label{fig:17}
\end{figure}

The adder circuit generates results as two outputs in the form of a sum and a carry, by adding two bytes applied to its input, referred to as a half-adder. Fig. \ref{fig:17} shows the half-adder symbol, while Tab. \ref{tab:1} shows the half-adder circuit truth table \cite{eki}.

\begin{table}
\caption{Half-adder circuit truth table }
\label{tab:1}
\begin{tabular}{|c|c|c|c|}
\hline
$A$&$B$&$S$&$C_0$\\
\hline
0&0&0&0\\
\hline
0&1&1&0\\
\hline
1&0&1&0\\
\hline
1&1&0&1\\
\hline
\end{tabular}
\end{table}
  
\begin{figure}
\begin{center}
\includegraphics[width=10.cm, height=7.cm]{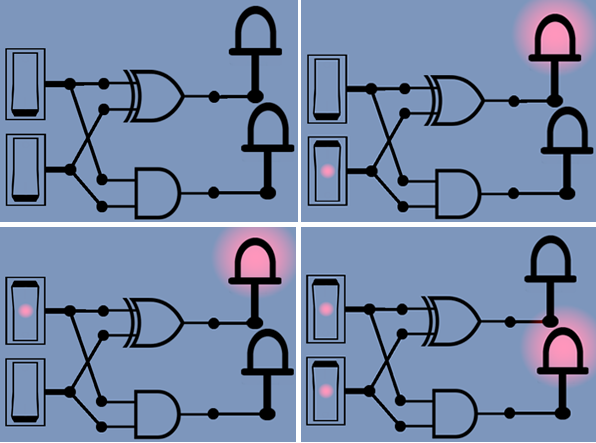}
\end{center}
\caption{Half-adder circuit screenshots}
\label{fig:18}
\end{figure}

Fig. \ref{fig:18} shows a screenshot of the simulation in which the inputs given in the truth table were applied in sequence. A Switch was used to apply the input loads. When the key was tapped, its location changed automatically. The Switch passed the Logic 1 load when the light was on, and the Logic 0 load when the light was off.

\subsection{Use of the Developed Mobile Application in Education }
\begin{figure}
\begin{center}
\graphicspath{{C:\Users\Ferda Nur\Desktop\LATEX\latex son\f19.png}}
\includegraphics[width=10.cm, height=6.cm]{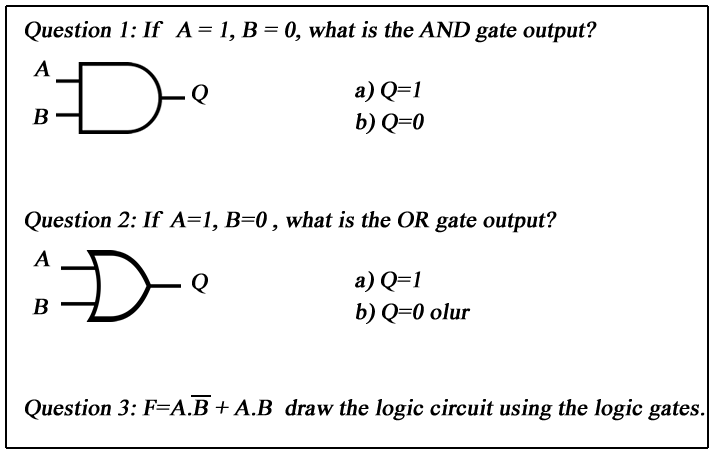}
\end{center}
\caption{Evaluation questions}
\label{fig:19}
\end{figure}

Through this educational instrument, an attempt was made to provide a better understanding of the running of logic gates and logic circuits through the use of simulations. The study carried out in March of the 2017–2018 academic year, involving two groups of students from a single class attending in the sixth grade of an Imam Hatip Secondary School affiliated to the Ministry of National Education. Each group contained 12 students who were assumed to be equivalent, and who had no prior knowledge of the subject. The first group attended a lecture about basic logic gates using only a blackboard, with no simulations demonstrated, while the second group, in addition to the lectures, were also shown a demonstration of the developed mobile application and its simulations. During the lesson, the software was run on a touch-screen laptop computer, and the lecture was delivered with presentations using a projector. At the same time, a tablet PC was circulated among the students on which they could perform the simulation themselves. Following the lecture, a written exam was applied to both groups. In the first and second questions of the exam, the input values of logic gates were given, and the resulting outputs were asked. In this section, both groups recorded the same level of success. In the third question, the students were requested asked to draw a logic circuit using logic gates. In the first group, 50 percent of the students were able to fully set up the circuit, 33 percent could only partially complete the task and 16 percent were not able to set it up at all. In the second group, 83 percent of the students set up the circuit in full, and 16 percent were only partially able to complete the task. The questions given to the students are presented in Fig. \ref{fig:19}. 

An evaluation of these results reveals that the students whose education was supported by the mobile application were more successful than the other group, indicating that the application is both useful and facilitating. It was also noted that the lecture delivered to the second group was more effective, and that the students gained a better understanding of the topic by being able to see and practice simulations.
\section{Conclusion and Suggestion}
In this study, an application that can run on both Android- and Windows-based mobile devices was developed that allows students attending classes such as Numerical/Digital Electronics, Logic Circuits, Basic Electronics Measurement and Electronic Systems in Turkeys Vocation and Technical Education Schools to easily carry out simulations of logic gates, as well as logic circuit tests performed using logic gates. The 2D- mobile application that runs on both Android and Windows platforms was developed using the C\# language on the Unity3D editor.

  To assess the usability of the mobile application, a one-hour training session was administered in March of the 2017–2018 academic year to two groups of students from the same class in the sixth grade of an Imam Hatip Secondary School affiliated to the Ministry of National Education. Each group contained 12 students who were assumed to be equivalent, and who had no prior knowledge of the subject. The first group attended a lecture on basic logic gates using only a blackboard, with no simulations demonstrated, while the second group, in addition to the lecture, was also given a demonstration of the developed mobile application and its simulations. Following the lectures, a written exam was applied to both groups.
  
  An evaluation of the results revealed that 83 percent of the students who had been given a demonstration of the mobile application were able complete the circuit during the exam, whereas in the other group, only 50 percent could complete the task. It was concluded from this that the application was both useful and facilitating for the students, and that the lecture delivered to the second group was more effective, as the students gained a better understanding of the topic by being able to see and practice simulations. 
  
  Using this simulation application is offers certain advantages, in that it is more secure; it allows, from an educational standpoint, the complexity of the learning process to be controlled; it can be used on both Android and Windows platforms; and it provides cost-savings by eliminating the unnecessary use of electronic components.
  
  The developed application can be presented to students undertaking Vocational and Technical Education, and can be improved further based on their feedback. Furthermore, the application can be made even more useful through the addition of other logic elements, such as flip-flops, multiplexers and demultiplexers and save processes. This work can also be made into a game and difficulty levels can be created from the circuit provided on the application screen. If the circuit is completed correctly by the user, both the level will increase and the points will be added. So the system will become educational and entertaining for students.


\end{document}